\documentstyle[aps,prl,preprint]{revtex}

\draft

\begin{document}
\author{Hong-shi Zong$^{1,2}$, Xiao-hua Wu$^{1,3}$, Xiao-fu L\"{u}$^{3}$, 
Chao-hsi Chang$^{2,4}$ and En-guang Zhao$^{2,4}$}
\address {$^{1}$ Department of Physics, Nanjing University, Nanjing 210093, P. R. China} 
\address{$^{2}$ CCAST(World Laboratory), P.O. Box 8730, Beijing 100080, P. R. China}
\address{$^{3}$  Department of Physics, Sichuan University,
Chengdu 610064, P. R. China}
\address{$^{4}$ Institute of Theoretical Physics, Academia Sinica, P.O. Box 2735, Beijing 100080, P. R. China}

\title{A New Approach to the Calculation of Some properties of Vacuum and $\pi$, $\sigma$ Mesons }
\maketitle

\begin{abstract}
It is shown in a systematic way that the instanton dilute liquid model can be combined with the global color symmetry model to form a new nonperturbative QCD model. Based on the quark propagator derived in the instanton dilute liquid approximation, the quark condensate $\langle\bar{q}q\rangle$, the mixed quark gluon condensate
$g_{s}\langle\bar{q}G_{\mu\nu}\sigma^{\mu\nu}q\rangle$, the four
quark condensate $\langle\bar{q} \Gamma q\bar{q} \Gamma q\rangle$  and tensor, pion vacuum susceptibilities  have been calculated at the mean field level
in the framework of the new nonperturbative QCD model. The numerical results are compatible with the values obtained within other
nonperturbative approaches. The calculated masses and decay constants of $\pi$ and $\sigma$ mesons are also close to the experimental values. 
\end{abstract}

\bigskip


\bigskip

E-mail: zonghs@chenwang.nju.edu.cn.

\bigskip

\begin{center}
{\bf \Large I. INTRODUCTION}
\end{center}

The non-perturbative structure of the QCD vacuum are characterized by 
the various condensates: such as the quark condensate $\langle\bar{q}q\rangle$, the mixed
quark gluon condensate $g_{s}\langle\bar{q}G_{\mu\nu}\sigma^{\mu\nu}q\rangle$, the four
quark condensate $\langle\bar{q} \Gamma q\bar{q} \Gamma q\rangle$, etc. These condensates
would vanish in a perturbative vacuum but do not vanish in the
non-perturbative QCD vacuum, and are of essential for describing the physics
of strong interaction[1,2]. Meanwhile susceptibilities of vacuum are also important quantities of strong interaction physics. They directly enter in the determination of hadron properties in the QCD sum rule external field approach[3--5]. In particular, tensor susceptibilities of the vacuum[6] are relevant for the determination of the tensor charge of the nucleon which is connected through deep-inelastic sum rules to the leading-twist nucleon transversity distribution[7]. The strong and parity--violating pion--nucleon coupling depends crucially upon $\chi^{\pi}$, the $\pi$ susceptibility[8]. In addition, as Goldstone bosons, the $\pi$ and $\sigma$ mesons provide ideal samples of exploring models of hadron structure in terms of the elementary degrees of freedom in quantum chromodynamics(QCD). They are then the ideal objects in investigating the nonperturbative aspect of QCD. Up to now, the main techniques for studying the nonperturbative aspects of QCD are lattice gauge theory[9], QCD sum rules[1--5], the instanton approximation[10-12] and some other phenomenological approaches. They have achieved quite success in describing different characteristics of low-energy QCD. However, each of these approach has its strengths and weaknesses. It is
interesting to combine two kinds of different nonperturbative QCD approaches to determine the vacuum condensates , vacuum susceptibilities and some properties of $\pi$ and $\sigma$ mesons within a unified model.

In the present paper, we try to address the above issues by combining the instanton dilute liquid model and the global color symmetry model(GCM) in a systematic way. Recently there are evidences that GCM provides a successful description of various nonperturbative aspect of strong interaction physics and the QCD vacuum as well as hadronic phenomena at low energies[13--15]. 

Although the instanton vacuum has been shown to provide a good phenomenological description of many hadronic properties[11,12], it has also some weaknesses[16]. To be concrete, it assert the vector self energy function of quark properator in instanton medium $A(q^2)$=1 which is not consistent with the corresponding one in Lattice QCD[17]. In addition, the $q^2 \gg 1 GeV^2$ behavior of scalar self energy function $B(q^2)$ of quark propagator in instanton medium disagrees with QCD. Therefore, how well does the instanton dilute liquid vacuum model the QCD vacuum is an interesting question. Furthermore, it is also an interesting question to what extent  can the inclusion of perturbative one gluon exchange term improve the predictions of the instanton model in a systematic way.

To proceed with our model one just explores its consequences, applying it to different systems , where possible, by comparisons with experimental data and other nonperturbative QCD approaches on their common domain. In this way we try to understand how well does our nonperturbative QCD model work. In order to keep this paper self-contained and easy to follow, a brief introduction of this nonperturbative QCD model is described in Sec.II. Some properties of the $\pi$ and $\sigma$ mesons, such as $f_{\pi}$, $m_{\pi}$, $f_{\sigma}$ and $m_{\sigma}$, will also be evaluated in Sec.II. From the determined generating functional, the vacuum condensates and vacuum susceptibilities are calculated in Sec.III. Finally
a brief discussion is given in Sec.IV.

\begin{center}
{\bf \Large II. A NONPERTURBATIVE QCD MODEL AND SOME PROPERTIES OF $\pi$ and $\sigma$}
\end{center}

It is well known that the QCD vacuum has topological charge because QCD is a nonabelian gauge theory. In order to consider the contribution of the nonperturbative QCD vacuum , instantons have to be introduced. It is believed that the instanton dilute liquid vacuum is a good approximation to the QCD vacuum configuration[18]. The perturbative contribution is deduced from the expansion about the instanton dilute liquid configuration, i.e., $A^{a}_{\mu}=A^{al}_{\mu}
+A^{aq}_{\mu}$, here $A^{al}_{\mu}$ are the fields of the instanton liquid and $A^{aq}_{\mu}$ are the quantum fluctuation fields of gluons. We will show in a reasonable way that the effect of the instanton liquid can be replaced by the Goldstone bosons. At the same time the hard perturbative term, in principle, can be calculated in this way. Thus there is no double counting in our approach.

Now we will use the path integral method to consider the contribution of the instanton dilute liquid. The generating functional for QCD in the Euclidean
metric is given by
\begin{equation}
Z[J, \bar{\eta}, \eta] = \int {\cal{D}}\bar{q} {\cal{D}} q 
{\cal{D}} A \exp\left\{-S 
+ \int d^4 x (\bar{\eta} q + \bar{q} \eta+J^{a}_{\mu}A^{a}_{\mu})\right\} \, ,
\end{equation}
where
$$ S = \int d ^4 x \left\{ \bar{q} \left[ \gamma _{\mu} \left(\partial_{\mu}
- i g \frac{\lambda ^{a}}{2} A^{a}_{\mu} \right) \right] q + \frac{1}{4}
G_{\mu \nu}^{a} G_{\mu \nu}^{a} \right\}, $$
and $G_{\mu \nu}^{a}=\partial_{\mu}A^{a}_{\nu}-\partial_{\nu}A^{a}_{\mu}+gf^{abc}A^{b}_{\mu}A^{c}_{\nu}$. We leave the gauge fixing term, the ghost field term and its integration measure to be understood. It is believed that the dilute liquid of instantons is a good approximation to the extreme of the pure gluon fields action[18]. The pure gluon fields action may be expanded about the dilute liquid of instantons in terms of the quantum fluctuation variables $A^{aq}_{\mu}$. With the functional expansion, the 
pure gluon fields action can be written as follows
\begin{eqnarray}
&&\int d ^4 x \frac{1}{4}
G_{\mu \nu}^{a}(x) G_{\mu \nu}^{a}(x) \nonumber \\
&&=\int d^4 x\frac{1}{4}
G_{\mu \nu}^{al}(x) G_{\mu \nu}^{al}(x)+\frac{1}{2}\int d^4 x d^4 y A^{aq}_{\mu}(x)
{\cal{D}}^{ab}_{\mu \nu}(x,y) A^{bq}_{\nu}(y)+\cdot\cdot\cdot, 
\end{eqnarray}
here $G_{\mu \nu}^{al}=\partial_{\mu}A^{al}_{\nu}-\partial_{\nu}A^{al}_{\mu}+gf^{abc}A^{bl}_{\mu}A^{cl}_{\nu}$, and the term ${\cal{D}}^{ab}_{\mu \nu}(x,y)$ can be regarded as the two point function of the quantum fluctuation fields. 

Substituting Eq.(2) into Eq.(1), the generating functional for QCD reads
\begin{eqnarray}
Z[J, \bar{\eta}, \eta] &=& \int {\cal{D}}\bar{q} {\cal{D}} q 
{\cal{D}} A^{q} \exp\left\{-S^{l} 
+ \int d^4 x \left[i g \bar{q}\frac{\lambda ^{a}}{2} A^{aq}_{\mu} q 
+ \bar{q} \eta+\bar{\eta} q
+J^{a}_{\mu}A^{al}_{\mu}+J^{a}_{\mu}A^{aq}_{\mu}\right]\right.\nonumber \\
&-&\frac{1}{2}\left.\int d^4 x d^4 y A^{aq}_{\mu}(x)
{\cal{D}}^{ab}_{\mu \nu}(x,y) A^{bq}_{\nu}(y)+\cdot\cdot\cdot\right\},
\end{eqnarray}
where
$$
S^{l} = \int d^4 x \left\{ \bar{q} \left[ \gamma _{\mu} \left(\partial_{\mu}
- i g \frac{\lambda ^{a}}{2} A^{al}_{\mu} \right) \right] q + \frac{1}{4}
G_{\mu \nu}^{al} G_{\mu \nu}^{al} \right\}.
$$

Using functional differentiation technique, Eq.(3) can be written as follows
\begin{eqnarray}
& &Z[J,\bar{\eta},\eta]\nonumber \\
&=& \int {\cal{D}} A^{q} \exp\left\{-\int d^{4}x(-\frac{\partial}{\partial\eta}ig\frac{\lambda^{a}}{2}\gamma_{\mu}A^{aq}_{\mu}\frac{\partial}{\partial\bar{\eta}}-J^{a}_{\mu}A^{aq}_{\mu})
-\frac{1}{2}\int d^4 x d^4 y A^{aq}_{\mu}(x)
{\cal{D}}^{ab}_{\mu \nu}(x,y) A^{bq}_{\nu}(y)+\cdot\cdot\cdot
\right\}
\nonumber\\
&&\times\int{\cal{D}}\bar{q}{\cal{D}}q \exp\left\{-S^l+\int d^4 x(\bar{\eta}q+\bar{q}\eta+J^{a}_{\mu}A^{al}_{\mu})\right\}.
\end{eqnarray}

Eq.(4) can be used to separates out the contribution of the instanton dilute liquid from the quantum fluctuation field. Thus we can study the effects of instanton dilute liquid and its quantum fluctuation fields separately. Now, we will pay attention the contribution of the instanton dilute liquid. In addition, it should be noted that it is interesting to consider further the effect of the perturbative gluon two-point function to the prediction of instanton model(or vice versa). 

Using again the functional differentiation technique, the generating functional for the contribution of the dilute liquid of instantons can be written as follows
\begin{eqnarray}
Z^{l}[J,\bar{\eta},\eta] &=& \int {\cal{D}} \bar{q} {\cal{D}} q\exp\left\{-S^l+\int d^4 x(\bar{\eta}q+\bar{q}\eta+J^{a}_{\mu}A^{al}_{\mu})\right\}
\nonumber\\
&=&\int {\cal{D}} \bar{q} {\cal{D}} q\exp\left\{\int d^4x(-\bar{q}\gamma_{\mu}\partial_{\mu} q+\bar{\eta}q+\bar{q}\eta+ig\bar{q}\frac{\lambda^{a}}{2}\gamma_{\mu} q\frac{\partial}{\partial J^{a}_{\mu}})\right\}\exp\left(W(J)\right),
\end{eqnarray}
where
$$
W[J]=\int d^4 x\left(-\frac{1}{4}G^{al}_{\mu\nu}G^{al}_{\mu\nu}+J^{a}_{\mu}A^{al}_{\mu}\right).
$$
It is easy to see that Eq.(5) describes the interaction between the quark and the dilute liquid of instanton.

It is obvious that $\exp(W(J))$ can be expanded in powers of $J^{a}_{\mu}$ as follows
\begin{equation}
\frac{exp (W[J])}{exp (W[J=0])}=\int d^4 x d^4 y\frac{1}{2}D^{ab}_{\mu\nu}(x,y) J^{a}_{\mu}(x)J^{b}_{\nu}(y)+W_{R}[J]
\end{equation}
where
$ W_{R} [ J_{\mu}^{a} ] = \sum_{n=3}^{\infty} \frac{1} {n !} \int d x_1
\cdots d x_n D_{\mu _1 \cdots \mu_n}^{a_1 \cdots a_n} (x_1,\cdots, x_n)
\Pi_{i=1}^{n} J_{\mu _i}^{a_i} (x_i) \, , $
and according to the model of the instanton dilute liquid, $D^{ab}_{\mu\nu}(x,y)=\langle 0\mid A^{al}_{\mu}(x)A^{bl}_{\nu}(y)\mid 0\rangle$ is evaluated on the vacuum configuration of the instanton dilute liquid.  $D^{ab}_{\mu\nu}$ is directly related to the gluon condensates.

Since $\exp(W[J=0])$ is a normalization factor, it can be omitted for convenience in the following. Introducing an auxiliary bilocal field $B^{\theta}(x,y)$ as in [13-15],
the generating functional of $Z^{l}[J,\eta,\bar{\eta}]$ can be written as
\begin{eqnarray}
Z^{l}[0,\bar{\eta}, \eta] & = & \exp {\left[W_{R} \left(ig \frac{\delta}{\delta
\eta(x)} \frac{\lambda^{a}}{2} \gamma_{\mu} \frac{\delta}{\delta
\bar{\eta}(x)} \right) \right] }  \nonumber \\  & & \times 
\int {\cal{D}}\bar{q} {\cal{D}} q {\cal{D}} B^{\theta}(x,y)
\exp\left\{-S [ \bar{q}, q, B^{\theta}(x,y)] + \int d^4 x ( \bar{\eta} q +\bar{q} \eta)\right\} \, , 
\end{eqnarray}
and 
$$ S[\bar{q}, q, B^{\theta}(x,y)] = \int \int d^4 x d^4 y \left[ \bar{q}(x)
G^{-1}(x,y;[B^{\theta}]) q(y) + \frac{B^{\theta}(x,y) B^{\theta}(y,x)}
{2 g^2 D(x-y)} \right] \; , $$
with
\begin{equation}
G^{-1}(x,y; [B^{\theta}]) = \gamma \cdot \partial \delta^{(4)}(x-y) +
\frac{1}{2} \Lambda^{\theta} B^{\theta}(x,y) \, ,
\end{equation}
where $\Lambda^{\theta} = K^{a} C^{b} F^{c}$ is determined by Fierz
transformation in color, flavor and Lorentz space. Since we have to take the average in the color space for calculation of the $g^2D^{ab}_{\mu\nu}(x-y)$, we have choosed a suitable gauge $ D_{\mu\nu}^{ab} (x-y)=\delta_{\mu\nu}\delta^{ab} D(x-y)$ in deriving the action $S[\bar{q}, q, B^{\theta}(x,y)]$.

Neglecting $W_{R}[J_{\mu}^{a}]$ as in [13-15] and performing the functional integral over ${\cal{D}} \bar{q}$ and
${\cal{D}} q$ in Eq. (7), we obtain
the modified GCM generating functional as
\begin{equation}
Z_{GCM}[\bar{\eta}, \eta] = \int {\cal{D}} B^{\theta}(x,y) \exp(-S [ \bar{\eta},
\eta, B^{\theta}(x,y)] ) \, , 
\end{equation}
where
\begin{eqnarray}
S[\bar{\eta}, \eta, B^{\theta}(x,y)] & = & - \mbox{Tr} \ln \left[\rlap/
\partial \delta (x-y) + \frac{1}{2} \Lambda^{\theta} B^{\theta}(x,y) \right]
\nonumber \\
& & + \int\int \left[\frac{B^{\theta}(x,y)
B^{\theta}(y,x)}{2 g^2 D(x-y)}
+ \bar{\eta}(x) G(x,y;[B^{\theta}])\eta(y) \right] .
\end{eqnarray}
From Eq.(10), it is easy to see that the $S[\bar{\eta}, \eta, B^{\theta}(x,y)]$ is the action of the bilocal bosonized fields. Here bilocal fields arise from the interaction between the quark fields and the dilute liquid of instanton.  

The saddle-point of the action is defined as
$\left.\delta S_[\bar{\eta}, \eta, B^{\theta}(x,y)]/\delta B^{\theta}(x,y)
\right\vert _{\eta =\bar{\eta}
= 0} =0$ and is given by
\begin{equation}
B^{\theta}_{0}(x-y)=g^2 D(x-y) tr_{\gamma C}[\Lambda^{\theta} G_{0}(x-y)],
\end{equation}
where $G_{0}$ stands for $G[B^{\theta}_{0}]$. These configurations are
related to vacuum condensates and
provide self-energy dressing of the quarks through the definition:
\begin{equation}
\Sigma(p)\equiv \frac{1}{2} \Lambda^{\theta}B^{\theta}_{0}(p)=\int d^4x e^{ip\cdot x}\left[\frac{1}{2} \Lambda^{\theta}B^{\theta}_{0}(x)\right] =i\gamma\cdot
p[A(p^2)-1]+B(p^2) , 
\end{equation}
where  $A(p^2)$ and $B(p^2)$ are the scalar and the vector self energy function of quark propagator, respectively. This dressing comprises the notion of $''$constituent quarks$''$ by providing 
a mass $M(p^2)=B(p^2)/A(p^2)$, reflecting a vacuum configuration with
dynamically broken chiral symmetry.

Recalling Eq.(8), at the mean field level(ie., the bilocal field $B^{\theta}_{q}(x,y)$ is simple replaced by the bilocal field vacuum configuration $B^{\theta}_{0}(x-y)$ in Eq.(8), physically this means that we only consider the influence of instanton medium to the quark propagator), we get
\begin{equation}
{\langle 0\vert T[q(x)\bar{q}(y)]\vert 0\rangle}^{-1}=\gamma\cdot\partial
\delta^{(4)}(x-y)+\frac{\Lambda^{\theta}}{2}B^{\theta}_{0}(x-y),
\end{equation}
with this equation the vacuum configuration of the bilocal field $B^{\theta}_{0}(x-y)$ is related to the quark propagator. It is well known that the propagation of light quarks was formulated quantitatively by treating quarks propagating in the instanton medium[12] and refined in its treatment of nonzero-mode propagation by Pobylitsa[19]. In momentum space, The quark propagator derived in the
instanton dilute liquid model can be written as[19]: 
\begin{equation}
G_{0}(q)=\frac{-i\gamma\cdot q+m(q^2)}{q^2+m^2(q^2)},~~~m(q^2) = \sqrt{\frac{\bar{\rho}}{2N_{c}}}\frac{q^2 \varphi^{2}(q)}{\sqrt{\int (2\pi)^{-4}d^{4}k k^2\varphi^{4}(k)}} \; ,
\end{equation}
where $\varphi(q) = \pi {\bar{R}}^2 \frac{d}{dz}[I_0(z) K_0(z)
-I_1(z) K_1(z)]$, in which $z= \frac{\vert q \vert \bar{R} }{2}$. 
$I_n(z) \; (K_n(z))$ \linebreak $(n=0,1)$ are the first (second) kind 
modified Bessel functions of order $n$. $\bar{\rho} \approx $
(200~MeV)$^4$ is the average density of the instantons. $\bar{R}= 
\frac{1}{3}~{\mbox{fm}}$ is the average radius of the instantons. 

As similar to the local field, the quantum bilocal fields $B^{\theta}_{q}(x,y)$ can also be expanded at the saddle point $B^{\theta}_{0}(x-y)$ and the fluctuations with respect to $B^{\theta}_{0}(x-y)$. Each fluctuation can be separated into two factors: the motion of the center of mass and the relative motion.
A separation of the internal and center-of-mass dynamics is achieved by considering the normal modes of the free kinetic operator of the bilocal fields in a manner that is analogous to the interaction representation of standard quantum field theory. As only the excitation of the Goldstone bosons is taken into account, we showed that the vacuum configuration
of the bilocal field $B(q^2)$ and the internal motion of the Goldstone bosons are equivalent to the dynamical mass $m(q^2)$ of quarks in the instanton dilute liquid approximation[20]. The mesons $\pi$ and $\sigma$ generated by the bilocal field in our model is then determined as
\begin{equation}
B^{\theta}_{q}(x,y)=B^{\theta}_{0}(x-y)+\sum_{\theta} \phi^{\theta}_{q}(\frac{x+y}{2})B(x-y),
\end{equation}
where the $\theta$ takes only $\pi$, $\sigma$ and $B(x-y)$ is the Fourier transformation of the $m(q^2)$. 
 
So far, we have finished the work of combining the instanton dilute liquid model and the GCM in a systematic way. As every parameter has been fixed in the instanton dilute liquid approximation, there are no any free parameter in our nonperturbative QCD model. It should be noted that strong
interaction phenomena provide evidences in favor of the instanton structure
of the QCD vacuum.  Lattice QCD[18,21] calculation shows that the instanton dilute liquid is a good approximation of QCD vacuum. In addition, we want to stress that the vacuum configuration of the bilocal field in GCM and the internal motion of the Goldstone bosons are directly related to the instanton dilute liquid approximation of the QCD vacuum. The results given below then reflected directly how well does the instanton dilute liquid vacuum model the QCD vacuum.

Based on Eqs.(10) and (15), the effective action of the Goldstone particles can be written as
\begin{eqnarray}
S[\sigma,\vec{\pi}]&=&-Tr ln\left[\gamma\cdot\partial\delta(x-y)+B(x-y)+[\frac{\sigma(\frac{x+y}{2})}{f_{\sigma}}+\frac{i}{f_{\pi}}\gamma_{5}\vec{\pi}(\frac{x+y}{2})\cdot \vec{\tau}]B(x-y)\right]\nonumber\\
&&+\frac{1}{2}\int d^4z\left[(1+\frac{\sigma(z)}{f_{\sigma}})^2+\frac{\vec{\pi}^2(z)}{f^2_{\pi}}\right]\int d^4w B(w)Tr[G_{0}(w)].
\end{eqnarray}

With Eq.(16), following the same way as shown in Refs.[13,20],  we obtain the normalization constants and the masses of $\pi$ and $\sigma$ mesons as
\begin{equation}
f^2_{\pi}=\frac{3}{8\pi^2}\int_{0}^{\infty} sds\left[\frac{3B^2+2B^3\frac{dB}{ds}+sB^3\frac{d^2B}{ds^2}+3sB^2(\frac{dB}{ds})^2}{(s+B^2)^2}-\frac{sB^2(1+2B\frac{dB}{ds})^2}{(s+B^2)^3}\right],
\end{equation}
\begin{equation}
m^2_{\pi}=\frac{3m}{2\pi^2 f^2_{\pi}}\int^{\infty}_{0}sds\frac{B(s)}{s+B^2(s)},
\end{equation}
\begin{eqnarray}
f^2_{\sigma}&=&\frac{3}{8\pi^2}\int^{\infty}_{0}sds\left\{\frac{\left[1+2B\frac{dB}{ds}+sB\frac{d^2B}{ds^2}-s(\frac{dB}{ds})^2\right]B^2}{(s+B^2)^2}+\right. \nonumber\\
&&\left.\frac{2\left[1+2B\frac{dB}{ds}+sB\frac{d^2B}{ds^2}+s(\frac{dB}{ds})^2\right]B^2(s-B^2)}{(s+B^2)^3}
-\frac{\left[1+2B(\frac{dB}{ds})^2\right]sB^2(s-B^2)}{(s+B^2)^4}\right\},
\end{eqnarray}
\begin{equation}
m^2_{\sigma}=\frac{3}{\pi^2f^2_{\sigma}}\int^{\infty}_{0}sds\frac{B^4(s)}{[s+B^2(s)]^2}.
\end{equation}

With Eqs.(14), (17), (19) and (20), we obtain the normalization constants $f_{\pi}$=86 MeV, $f_{\sigma}$=77 MeV and $m_{\sigma}$=649 MeV. As the $\pi$ normalization constant can be simply related to the decay constant of $\pi$ when the vector self-energy function of quark propagator $A(q^2)=1$[22], the calculated result of $f_{\pi}$ is quite close to the experimental data $f_{\pi}$=93 MeV. Meanwhile the obtained  $m_{\sigma}$ is quite reasonable which is in the scope of experiments(500--700) MeV[23]. Furthermore, the expression of $m_{\pi}$
reproduces a PCAC result that $m_{\pi}\not= 0$ only if the current quark mass $m\not= 0$. This agrees well with the Goldstone-Boson feature of the pion. Moreover, we fit $m_{\pi}=139$ MeV as the current quark mass $m$ is taken to be 3.7 MeV.

From Eq.(18) and the calculation of the two quark condensate $\langle\bar{q}q\rangle$ below, we have
\[
f^2_{\pi}m^2_{\pi}=-2m\langle\bar{q}q\rangle.
\]
which is the current algebra result derived by Gell--Mann, Oakes and Renner(GMOR).

\begin{center}
{\bf \Large III. CALCULATION OF SOME PROPERTIES OF VACUUM}
\end{center}
\begin{center}
{\bf \large A. EVALUATION OF VACUUM CONDENSATES}
\end{center}

Once the generating functional in our model are determined,
one can calculate the $\langle\bar{q}q\rangle$,
$g_{s}\langle\bar{q}G_{\mu\nu}\sigma^{\mu\nu}q\rangle$ and
$\langle\bar{q} \Gamma q\bar{q} \Gamma q\rangle$ vacuum condensate at
the mean field level. 
Previous studies of these condensates
include QCD sum rules where it was treated as fitted parameter in the
analysis of various hadron channels [1,2], quenched lattice QCD[24], the
instanton liquid model[25] and the model of confining gluon propagator used in the Schwinger--Dyson form for GCM[26].

With the modified GCM generating functional, it is straightforward
to calculate the vacuum expectation value of any quark operator of the form
\begin{equation}
O_{n}\equiv(\bar{q}_{j1}\Lambda^{(1)}_{j1i1}q_{i1})
(\bar{q}_{j2}\Lambda^{(2)}_{j2i2}q_{i2})\cdots
(\bar{q}_{jn}\Lambda^{(n)}_{jnin}q_{in}),
\end{equation}
in the mean field vacuum. Here the $\Lambda^{(i)}$ stands for an operator in
Dirac, flavor and color space.

Take the appropriate number of derivatives with respect to external sources
$\eta_{i}$ and $\bar{\eta_{j}}$ of Eq.(9)
and putting $\eta_{i}=\bar{\eta}_{j}=0$[27],
one obtains
\begin{equation}
\langle:O_{n}:\rangle=(-)^{n}\sum_{p}[(-)^{p}\Lambda^{(1)}_{j1i1}
\cdots\Lambda^{(n)}_{jnin}(G_0)_{i1jp(1)}\cdots(G_0)_{injp(n)}],
\end{equation}
where p stands for a permutation of the n indices. In particular we obtain
the nonlocal quark condensate $\langle:\bar{q}(x)q(0):\rangle$ as
\begin{eqnarray}
\langle:\bar{q}(x)q(0):\rangle_{\mu}&=&(-)tr_{\gamma C}[G_{0}(x,0)]\nonumber\\
&=&(-4N_{c})\int^{\mu}_{0}\frac{d^4 p}{(2\pi)^4}\frac{B(p^2)}
{p^2A^2(p^2)+B^2(p^2)}e^{ipx}\nonumber\\
&=&(-)\frac{12}{16\pi^2}\int^{\mu}_{0}ds s\frac{B(s)}
{sA^2(s)+B^2(s)}[2\frac{J_1(\sqrt{sx^2})}{\sqrt{sx^2}}],
\end{eqnarray}
where $\mu$ is the cutoff which we choose to be 1 $GeV^2$
as in Refs.[26] and [28].
It is natural to introduce a cutoff because our model is an effective theory and one expects that there will be a characteristic scale in which the theory is effective. This situation is very similar to the determination of vacuum condensates in Nambu and Jona-Lasinio(NJL) model.

At x=0 the expression for the local condensate $\langle:\bar{q}q:\rangle$ is recovered:
\begin{equation}
\langle:\bar{q}q:\rangle_{\mu}=(-)tr_{\gamma C}[G_{0}(x,0)]|_{x=0}=(-)\frac{3}{4\pi^2}\int^{\mu}_{0}ds s
\frac{B(s)}{sA^2(s)+B^2(s)}.
\end{equation}
The nonlocality $g(x^2)$ can be obtained immediately by dividing Eqs.(23)by Eq.(24).
In Table I, we display the nonlocal quark condensate $g(x^2)$ in the text. By comparing it with the corresponding results mentioned in Ref.[29], we demonstrate that the nonlocal quark condensate is very robust with respect to using our approach or the models of confining gluon propagators.

\begin{tabular}{lccccccccccr}
\multicolumn{12}{c}{Table. I. The nonlocal quark condensate
$g(x)=\langle:\bar{q}(x)q(0):\rangle/\langle:\bar{q}(0)q(0):\rangle$.}\\ \hline\hline
$x^2(GeV^{-2})$ &0.0 &2.0 &4.0 &6.0 &8.0 &10  &12  &14  &16  &18  &20 \\ \hline
$g(x^2)$     &1.0 &0.89&0.79&0.71&0.63&0.55&0.50&0.44&0.38&0.34&0.30\\ \hline\hline  \end{tabular}

\bigskip

Another important vacuum condensate is the nonlocal four quark
condensate in the mean filed vacuum. From Eq.(22), one has
\begin{eqnarray}
& &\langle:\bar{q}(x)\Lambda^{(1)}q(x)\bar{q}(y)
\Lambda^{(2)}q(y):\rangle_{\mu}\nonumber\\
&=&-tr_{\gamma C}[G_{0}(y,x)\Lambda^{(1)}
G_{0}(x,y)\Lambda^{(2)}]
+tr_{\gamma C}[G_0(x,x)\Lambda^{(1)}]
tr_{\gamma C}[G_{0}(y,y)\Lambda^{(2)}].
\end{eqnarray}

By means of Eq.(25), one can calculate all kinds of nonlocal four quark condensates at the mean field level. For instance in case of $\Lambda^{(1)}$
=$\Lambda^{(2)}$=$\gamma_{\mu}\frac{\lambda^a_{C}}{2}$ one find from Eq.(25):
\begin{eqnarray}
& &\langle:\bar{q}(x)\gamma_{\mu}\frac{\lambda^a_{C}}{2}q(x)\bar{q}(0)
\gamma_{\mu}\frac{\lambda^a_{C}}{2}q(0):\rangle_{\mu}\nonumber\\
&=&(-)\int^{\mu}_{0}\int^{\mu}_{0}
\frac{d^4p}{(2\pi)^4}\frac{d^4q}{(2\pi)^4}e^{ix\cdot(p-q)}
\left[4^3\frac{B(p^2)}{A^2(p^2)p^2+B^2(p^2)}\frac{B(q^2)}
{A^2(q^2)q^2+B^2(q^2)}\right.\nonumber\\
& &\left.+ 2\times 4^2\frac{A(p^2)}{A^2(p^2)p^2+B^2(p^2)}\frac{A(q^2)}
{A^2(q^2)q^2+B^2(q^2)}p\cdot q\right].
\end{eqnarray}
Similarly, at x=0 the expression for the local four quark condensate
$\langle:\bar{q}\gamma_{\mu}\frac{\lambda^a_{C}}{2}q\bar{q}
\gamma_{\mu}\frac{\lambda^a_{C}}{2}q:\rangle$ is recovered:
\begin{equation}
\langle:\bar{q}\gamma_{\mu}\frac{\lambda^a_{C}}{2}q\bar{q}
\gamma_{\mu}\frac{\lambda^a_{C}}{2}q:\rangle_{\mu}=(-4^3)[\int^{\mu}_{0}\frac{d^4p}{(2\pi)^4}
\frac{B(p^2)}{A^2(p^2)p^2+B^2(p^2)}]^2=(-)\frac{4}{9}\langle:\bar{q}q:\rangle^2
,
\end{equation}
i.e. for the local four quark condensate, our result is consistent
with the vacuum saturation assumption of Ref.[1]. However, if one considers the
nonlocal four quark condensate, it should be noted that the contribution of
the second term of the right-hand of Eq.(26) is about a 30\% correction and can not be neglected[30]. 

The introduction of
nonlocal condensates[31] have been shown to be useful for representing the bilocal vacuum
matrix elements needed for the pion wave function[32] and pion form factor[33] 
for low and medium momentum transfer. In the external field method of QCD sum rules, one
does not carry out OPE for the vacuum matrix element of the bilocal operators but introduce 
new phenomenological parameters to characterize the space-time structure of the
nonlocal condensates. Previous studies of calculating this nonlocal condensate include 
the instanton liquid model[34] and the model of confining gluon propagator used in the Schwinger--Dyson form for GCM[29].
Although $\langle:\bar{q}(x)q(0):\rangle$ is not gauge invariant, it is useful for describing 
the scalar part of nonperturbative quark propagator.
Therefore, we use our modified version of GCM to calculate the $\langle :\bar{q}(x)q(0):\rangle$ and nonlocal four quark condensate. 

As far as the mixed condensate
$g_{s}\langle\bar{q}G_{\mu\nu}\sigma^{\mu\nu}q\rangle$, one
can use the method described by Refs.[26] to obtain the mixed condensate in
Euclidean space
\begin{equation}
g_{s}\langle\bar{q}G_{\mu\nu}\sigma^{\mu\nu}q\rangle_{\mu}
=(-)(\frac{N_{c}}{16\pi^2})\{\frac{27}{4}
\int_{0}^{\mu}dss\frac{B[2A(A-1)s+B^2]}
{A^2s+B^2}+12\int_{0}^{\mu}dss^2\frac{B(2-A)}{A^2s+B^2}\}
\end{equation}

In Table II. we display the results for $\langle\bar{q}q\rangle$ and
$g_{s}\langle\bar{q}G_{\mu\nu}\sigma^{\mu\nu}q\rangle$ in our approach and compare it
with the corresponding values which were obtained by other nonperturbative approaches:
QCD sum rules[3], quenched lattice QCD[24], the instanton liquid model[25]
and the model of confining gluon propagator used in the Schwinger--Dyson form for GCM[26].

\begin{tabular}{llr} 
\multicolumn{3}{c}{Table. II. $\langle\bar{q}q\rangle$ and $\langle g_{s}\bar{q}G_{\mu\nu}\sigma^{\mu\nu}q\rangle$
in different non-perturbative approaches}\\ \hline\hline            
                 & (--)$\langle\bar{q}q\rangle^{\frac{1}{3}}$ & (--)$\langle g_{s}\bar{q}\sigma Gq\rangle^{\frac{1}{5}}$ \\ \hline
this work      & 217 MeV                     & 429 MeV              \\
QCD sum rules[3] & 210 --230 MeV                 & 375 --395 MeV  \\
quenched lattice[24]& 225 MeV                   & 402 --429 MeV   \\
instanton liquid model[25] & 272 MeV              & 490 MeV \\
the model of confining gluon propagator in GCM[26] & 150 --180 MeV & 400 --460 MeV \\ \hline\hline
\end{tabular}

\bigskip

Table II. shows that our results for $\langle\bar{q}q\rangle$ and
$g_{s}\langle\bar{q}G_{\mu\nu}\sigma^{\mu\nu}q\rangle$ are compatible with the values
obtained within other nonperturbative methods, especially in QCD sum
rules[3] and quenched lattice QCD[24]. 

\begin{center}
{\bf \large B. EVALUATION OF VACUUM SUSCEPTIBILITIES}
\end{center}

Now we pass to the calculation of pion and tensor vacuum susceptibilities.
Quite different values of tensor vacuum susceptibility have been reported in literature[35-38]. Therefore it is interesting to address this issue in our model. 

In the external field of QCD sum rule two--point method, one often encounters
the quark propagator in the presence of the
$J^{\Gamma}(y)=\bar{q}(y)\Gamma q(y)$ current($\Gamma$ stands for the appropriate
combination of Dirac and flavor matrices). 
\begin{eqnarray}
S^{cc'\Gamma}_{\alpha\beta}(x)&=&\langle 0|T[q^{c}_{\alpha}(x)\bar{q}^{c'}_{\beta}(o)]0
\rangle_{J^{\Gamma}}\nonumber \\
&=&S^{cc'\Gamma,PT}_{\alpha\beta}(x)
+S^{cc'\Gamma,NP}_{\alpha\beta}(x),
\end{eqnarray}
where $S^{\Gamma,PT}_{q}(x)$ is the quark propagator coupled perturbatively
to the current and $S^{\Gamma,NP}_{q}(x)$ is the nonperturbative quark
propagator in the presence of the external current $J^{\Gamma}$(one should
note that the external current should be taken to be $J^{\Gamma}\phi_{\Gamma}$, where $\phi_{\Gamma}$
is the value of the external field. In what follows, to simplify the notation, we will take the $\phi_{\Gamma}$=1, which does not affect our results).
The vacuum susceptibility $\chi^{\Gamma}$ in the QCD sum rule two--point  external field treatment can be defined as[39]
\begin{eqnarray}
S^{cc'\Gamma,NP}_{\alpha\beta}(x)&=&\langle 0|:q^{c}_{\alpha}(x)
\bar{q}^{c'}_{\beta}(0):|0\rangle_{J^{\Gamma}}\nonumber\\
&=&-\frac{1}{12}\Gamma_{\alpha\beta}\delta_{cc'}
\chi^{\Gamma}H(x)\langle 0|:\bar{q}(0)q(0):|0\rangle ,
\end{eqnarray}
where the phenomenological function $H(x)$ represents the nonlocality of the
two quark nonlocal condensate. Note that H(0)=1.

The presence of external field implies that $S^{cc'\Gamma}_{\alpha\beta}(x)$
is evaluated with an additional term $\Delta L\equiv -J^{\Gamma}\cdot\phi_{\Gamma}$
added to the usual QCD Lagrangian. In three point method of QCD sum rule, if one takes only a linear external
field approximation, the $S^{cc'\Gamma,NP}_{\alpha\beta}(x)$ in Euclidean space
is given by
\begin{equation}
S^{cc'\Gamma,NP}_{\alpha\beta}(x)=
\int d^{4}y~e^{-iq\cdot y}\langle 0|:q^{c}_{\alpha}(x)\bar{q}^{e}(y)
\Gamma q^{e}(y)\bar{q}^{c'}_{\beta}(0):|0\rangle .
\end{equation}

Using Eqs.(30), Eq.(31) and note that $H(0)=1$ we have
\begin{equation}
-\frac{1}{12}\delta_{cc'}\Gamma_{\alpha\beta}\chi^{\Gamma}
\langle 0|:\bar{q}(0) q(0):|0 \rangle
=\int d^{4}y~e^{-iq\cdot y}\langle 0|:q^{c}_{\alpha}(0)\bar{q}^{e}(y)
\Gamma q^{e}(y)\bar{q}^{c'}_{\beta}(0):|0\rangle .
\end{equation}

Multiplying Eq.(32) by $\Gamma_{\beta\alpha}\delta_{cc'}$, we get
\begin{equation}
\chi^{\Gamma}a
=-\frac{16\pi^2}{tr_{\gamma}(\Gamma\Gamma)}\int d^{4}y~ e^{-iq\cdot y}
\langle 0|:\bar{q}^{c}(0)\Gamma q^{c}(0)
\bar{q}^{e}(y)\Gamma q^{e}(y):|0\rangle ,
\end{equation}
with $a$=-$(2\pi)^2\langle 0|:\bar{q}(0) q(0):|0 \rangle$. Eq.(33) shows that the vacuum susceptibilities originates from the nonlocal four quark condensate contribution. This conclusion is the same as that of Ref.[38] which addressed the problem from a completely different viewpoint, using the concept of duality(more details can be found in Ref.[38]).

In the case of tensor current($\Gamma=\sigma_{\mu\nu}$), from Eqs.(26), Eq.(33) and the fact $tr_{\gamma}(\sigma_{\mu\nu}\sigma_{\mu\nu})=48$, we have the tensor vacuum susceptibility $\chi^{T}a$:
\begin{eqnarray}
&&\chi^{T}a \nonumber\\
&=&Lim_{q\rightarrow 0}(\pi^2)\int d^4y~e^{-iq\cdot y}tr_{\gamma}\left[
\int \frac{d^4p}{(2\pi)^4}\frac{-i\gamma\cdot p A+B(p^2)}{A^2p^2+B^2(p^2)}
e^{-ip\cdot y}\sigma_{\mu\nu}
\int \frac{d^4l}{(2\pi)^4}\frac{-i\gamma\cdot l A+B(q^2)}{A^2l^2+B^2(l^2)}
e^{il\cdot y}\sigma_{\mu\nu}\right] \nonumber\\
&=&Lim_{q\rightarrow 0}3\int^{\mu}_{0}sds\left[\frac{B(s)}{A^2s+B^2(s)}\right]^2\simeq~0.23~GeV^2 ,
\end{eqnarray}
with $\chi^{T}a$ an opposite sign and a factor $(4\times\pi^2)$ larger than the definition in Refs.[35,36,38]. In order to compare it with the estimation in Refs.[35,36,38],we have
\begin{equation}
\frac{\chi^{T}a}{(-4\times\pi^2)}\simeq~-0.0058~ GeV^2 ,
\end{equation}
which is very close to the result  -0.0055 $GeV^2$ 
obtained in Ref.[38]. Notice, that the result of present paper is less than the estimations -0.010 $\sim$ -0.008 $GeV^2$ obtained in Ref.[36], and has opposite sign and is several times larger than the previous estimation in Ref.[35]. In order to make it easy to compare our result with previous estimations of vacuum tensor susceptibilities, a simple table is listed below
\begin{center}
\begin{tabular}{ccccc}
\multicolumn{5}{c}{Table.III. Vacuum tensor susceptibilities($GeV^2$).}\\ \hline\hline  Ref[35]  & Ref[36]  & Ref[37] & Ref[38] & this work\\
 +0.002&-0.008  &+0.009$\leftrightarrow$+0.017    &-0.0055  & -0.0058\\ \hline\hline
\end{tabular}
\end{center} 

Now we turn to the calculation of Pion vacuum susceptibility(The Pion vacuum susceptibility is crucial to the strong and party-violating pion-nucleon coupling). In the case of pseudoscalar current, from Eqs.(26) and (33), we have the Pion vacuum susceptibility $\chi^{\pi}a$:
\begin{eqnarray}
&&\chi^{\pi}a \nonumber\\
&=&12\pi^2\int d^4y~tr_{\gamma}\left[
\int \frac{d^4p}{(2\pi)^4}\frac{-i\gamma\cdot p A+B(p^2)}{A^2p^2+B^2(p^2)}
e^{-ip\cdot y}\gamma_{5}
\int \frac{d^4q}{(2\pi)^4}\frac{-i\gamma\cdot q A+B(q^2)}{A^2q^2+B^2(q^2)}
e^{iq\cdot y}\gamma_{5}\right] \nonumber\\
&=&3\int^{\mu}_{0}sds\frac{1}{A^2s+B^2(s)}=2.56~GeV^2,
\end{eqnarray}
which is compatible with the range $\chi^{\pi}a\simeq (1.7-3.0)~GeV^2$ obtained within a phenomenological approach[39]. The integral in Eq.(36) is divergent, as is inherent in effective models, hence a cutoff is introduced in the integral of Eq.(33) as in the Nambu and Jona-Lasinio model.

\begin{center}
{\bf \Large IV. Discussion}
\end{center}
In the present paper, we first gave a systematic way to combine two kinds of different nonperturbative QCD approaches, ie., the instanton dilute liquid model and the GCM, then based on the quark propagator derived in the instanton dilute liquid approximation, we have
determined the quark condensate $\langle\bar{q}q\rangle$, the mixed quark gluon condensate
$g_{s}\langle\bar{q}G_{\mu\nu}\sigma^{\mu\nu}q\rangle$, the four
quark condensate $\langle\bar{q} \Gamma q\bar{q} \Gamma q\rangle$  and tensor, pion vacuum susceptibilities at the mean field level
in the framework of our nonperturbative QCD model which used the quark propagator derived in instanton dilute liquid vacuum as input. The numerical calculation
shows our results is compatible with the range obtained within other
nonperturbative approaches. In particular, the numerical calculation of the tensor vacuum susceptibility
shows that our results is very close to the one obtained in Ref.[38] and has opposite sign to that of Refs.[35] and [37]. As has been pointed out in Ref.[35] experimental measurement of the  tensor charge of the nucleon is possible, and one might be able thereby to test the theoretical predictions of the tensor susceptibility in the future.
Meanwhile we have also calculated the masses and decay constants of $\pi$ and $\sigma$ mesons. The calculated results agree with experimental data

It should be noted that the cutoff must be introduced in the calculation of the vacuum condensates and susceptibilities. This is consistent with the philosophy of the model: it is applicable below the chiral symmetry breaking scale(about 1 $GeV$). Other nonperturbative QCD models, such as the NJL model and chiral quark model has the same limitation. Some quantities, such as the mass and normalization constants of $\pi$ and $\sigma$, are not sensitive to this cutoff. For example, if one take the same cutoff $\mu=1 GeV^2$ in the integral of Eq.(17) and Eq.(19), one get $f_{\pi}$= 85 MeV and $f_{\sigma}$= 76 MeV which is scarcely affected by the cutoff. This is because every quantum fluctuation in GCM can be separated into the motion of the center of mass and the relative motion. The relative motion factor(form factor) plays the role of suppressing the large momentum behavior in the integral. However if one calculate the vacuum properties(such as vacuum condensates and susceptibilities), there are no such form factor to suppress the large momentum in the integral(some integral such as Eq.(36) is divergent). hence a cutoff is necessary in calculating the vacuum properties.

Finally, we like to emphasize that there are no any
free parameter in our model, the only input is the quark propagator in instanton vacuum. With this model, the calculated results of
masses and the decay constants of the mesons $\pi$ and $\sigma$ and
some vacuum properties agree with experimental or experiential dates quite well. We conclude that quark propagator in instanton medium contain valuable information of nonperturbative QCD and is consistent with recent models of confining gluon propagators used in the Schwinger--Dyson formalism for the global color symmetry model.

\noindent{\large \bf Acknowledgement}

This work was supported in part by the National Natural Science Foundation
of China. 

\vspace*{2.cm}
\noindent{\large \bf References}
\begin{description}
\item{[1]} M. Shifman, A. Vainshtein and V. Zakharov, Nucl. Phys 
{\bf B147}, 385 (1979).
\item{[2]}  L. Reinders, H. Rubinstein and S. Yazaki, Phys. Rep.
{\bf 127}, 1 (1985); S. Narison, QCD Spectral Sum Rules (World Scientific, Singapore
, 1989), and references therein.
\item{[3]} B. L. Ioffe and A. V. Smilga, Nucl. Phys. {\bf B232}, 109 (1984).
\item{[4]} I. I. Balitsky and A. V. Yung, Phys. Lett. {\bf B129}, 328 (1983).
\item{[5]} S. V. Mikhailov and A. V. Radyushkin, JETP Lett. {\bf 43},
712 (1986); Phys. Rev. {\bf D45}, 1754 (1992); A. P. Bakulev and A. V. Radyushkin, Phys.  Lett. {\bf B271}, 223 (1991).
\item{[6]} H. He and X. Ji, Phys. Rev. {\bf D52}, 2960 (1995).
\item{[7]} J. Ralston and D. E. Soper, Nucl. Phys 
{\bf B152}, 527 (1979); R. L. Jaffe and X. Ji, Phys. Rev. Lett{\bf 67}, 552 (1991).
\item{[8]} E. M. Henley, W. -Y. Hwang and L. S. Kisslinger, Phys. Lett. {\bf B367}, 21 (1996).
\item{[9]} K. G. Wilson, in New Phenomena in Subnuclear Physics, edited by A. Zichichi(Plenum, New York,1977); I. Montavey and G. M\"{u}nster, Quantum Fields on a Lattice(Cambridge University Press, Combridge, England, 1994).
\item{[10]} C. G. Callan, Jr., R. Dashen and D, J. Gross, Phys. Rev. {\bf D17}, 2717 (1978).
\item{[11]} E.~V. Shuryak, Nucl. Phys. {\bf B203}, 93 (1982); Phys. Rep. {\bf 115}, 151 (1984); T. Schafer and
E. V. Shuryak. Rev. Mod. Phys. {\bf 70}, 323 (1998).
\item{[12]} D.~I. Dyakonov and X. Yu Petrov, Nucl. Phys. {\bf B245}, 259 (1984); {\bf B272}, 457 (1986). 
\item{[13]} R. T. Cahill and C. D. Roberts, Phys. Rev. {\bf D32}, 2419 (1985).
\item{[14]} M. R. Frank and T. Meissner, Phys. Rev {\bf C53}, 2410 (1996); M. R. Frank and C. D. Roberts, Phys. Rev {\bf C53}, 390 (1996).
\item{[15]} C. D. Roberts and A. G. Williams, Prog. Part. Nucl. Phys.{\bf 33},
477 (1994), P. C. Tandy, Prog. Part. Nucl. Phys. 39, 117 (1997) and references therein.
\item{[16]} L. S. Kisslinger, M. Aw, A. Harey and O. Linsuain, Phys. Rev. {\bf C60}, 065204 (1999).
\item{[17]} J. -T. Skullerud and A. G. Williams, hep-lat/9909142.
\item{[18]} J. W. Negele, Nucl. Phys. {\bf A670}, 14c (2000).
\item{[19]} P. V. Pobylitsa, Phys. Lett. {\bf B226}, 387 (1989).
\item{[20]} Xiao-fu L\"{u}, Yu-xin Liu, Hong-shi Zong and En-guang Zhao,
Phys. Rev. {\bf C58}, 1195 (1998).
\item{[21]} M. C. Chu, J. M. Grandy, S. Huang and J. Negele, Phys. Rev.
{\bf D49}, 6039 (1994); J. Negele, ``Lattice QCD'' presented at the
International Summer School and Workshop on Nuclear QCD, Beijing,1995.
\item{[22]} P. maris, C. D. Roberts and P. C. tandy, Phys. Lett. {\bf B420}, 267 (1998).
\item{[23]} Particle Data Group, Phys Rev. {\bf D54}, 1 (1996) and references therein.
\item{[24]} M. Kremer and G. Schierholz,  Phys. Lett. {\bf B194}, 283 (1987).
\item{[25]} M. V. Polyakov and C. Weiss, Phys. Lett. {\bf B387}, 841 (1996). 
\item{[26]} T. Meissner, Phys. Lett. {\bf B405}, 8 (1997).
\item{[27]} J. Negele and H. Orland, Quantum Many-particles Systems
(Addison-Wesley 1988).
\item{[28]} Hong-shi Zong, Xiao-fu L\"{u}, Jian-zhong Gu, Chao-hsi Chang and En-guang Zhao,
Phys. Rev. {\bf C60}, 055208 (1999).
\item{[29]} L. S. Kisslinger and T. Meissner, Phys. Rev. {\bf C57}, 1528 (1998).
\item{[30]} Hong-shi Zong, Xiao-fu L\"{u}, En-guang Zhao and Fan Wang, Commun Theor. Phys.(Beijing, China) {\bf 33}, 687 (2000); Hong-shi Zong, Xiao-fu L\"{u}, Fan Wang, Chao-hsi Chang and En-guang Zhao, Commun Theor. Phys.(Beijing, China) {\bf 34}, 563 (2000).
\item{[31]} S.V.Mikhailov and A. V. Radyushkin, Sov. J. Nucl. Phys. {\bf 49} (1989) 494;
\item{[32]} S.V.Mikhailov and A. V. Radyushkin, Phys. Rev. {\bf D45} (1992) 1754.
\item{[33]} A. P. Bakulev and A. V. Radyushkin, Phys. Lett. {\bf B 271} (1991) 223.
\item{[34]} A. E. Dorokhov, S. V. Esaibegyan, S.V.Mikhailov and A. V. Radyushkin, Phys. Rev. {\bf D56} (1997) 4065; A. E. Dorokhov and L. Tomio
hep-ph/9803329; A. E. Dorokhov, S. V. Esaibegyan, A. E. Maximov, and
S.V.Mikhailov, Eur. Phys. J. {\bf C13}, 331 (2000).
\item{[35]} H. He and X. Ji, Phys. Rev. {\bf D54}, 6897 (1996) 
\item{[36]} V. M. Belyaev and A. Oganesian, Phys. Lett.{\bf B395},307 (1997).
\item{[37]} L. S. Kisslinger, Phys. Rev. {\bf C59}, 3377 (1999).
\item{[38]} A. P. Bakulev and S. V. Mikhailov, Eur. Phys.J.{\bf C17},129 (2000).
\item{[39]} M. B. Johnson and L. S. Kisslinger, Phys. Rev. {\bf D57},2847 (1998).

\end{description}

\end{document}